\documentstyle[prb,twocolumn,aps,graphicx]{revtex}
\tighten
\begin{document}
\draft


\wideabs{
\title{Progressive suppression of spin relaxation in 2D channels of 
finite width}
\author{A. A. Kiselev and K. W. Kim}
\address{Department of Electrical and Computer Engineering,
North Carolina State University, Raleigh, NC 27695-7911}

\maketitle
\begin{abstract}
We have investigated spatio-temporal kinetics of electron spin polarization
in semiconductor narrow 2D strip and explored the ability to
manipulate spin relaxation.
Information about spin of the conduction electrons and mechanisms of
spin rotation is incorporated into transport Monte Carlo simulation
program. A model problem, involving linear-in-$k$ splitting of the
conduction band, responsible for the D'yakonov-Perel' mechanism of spin
relaxation in the zinc-blende semiconductors and heterostructures, is
solved numerically to yield the decay of spin polarization of an ensemble of
electrons in the 2D channel of finite width. For very wide channels, a
conventional 2D value of spin relaxation is obtained. With decreasing
channel width the relaxation time soares rapidly by orders of
magnitude. Surprisingly, the cross-over point between 2D and quasi-1D
behavior is found to be at tens of electron mean-free paths. Thus,
classically wide channels can effectively suppress electron spin
relaxation.
\end{abstract}
}

\section{Introduction}
\label{introduction}

Spintronics,\cite{prinz,gregg} a nontrivial extention of the
conventional electronics, adds functionality utilizing the carrier spin
degree of freedom. Spin can potentially be used as a by far more
capacious quantum information storage cell, be involved in the transfer
of information, for elaborate schemes of information processing, both
quantum and classical, and be integrated with electric-charge
counterparts in combined designs. Electron or nuclear spin manipulation
is believed to be the key component of the potential realizations of
a quantum computer (for example, see Ref.~\onlinecite{kane}).

Several devices were proposed so far including spin
hybrid\cite{monsma}
and field effect transistors,\cite{datta}
tunneling structures with
magnetic layers,\cite{moodera} and spin-based memory.\cite{dax}
The most notable success was obtained with giant magnetoresistance (GMR)
effect devices that rely on the
variation of electron scattering in a multilayer stack of ferromagnetic
films separated by nonmagnetic materials.\cite{ansermet,allen} Though the
variation of the current through the structure is small,
it is sufficient for detection, and the excellent sensitivity of the
device to weak external magnetic fields (typically $1\%$ change of
resistance per oersted) opened the door for massive
data storage applications.
Spin coherence is a pivotal prerequisite for the operability of the
prospective spintronic devices.

Concerning experimental realization of spintronic elements,
most propositions rely on the injection of spin-polarized electrons
from the ferromagnetic layer and suffer from the poor quality of the
ferromagnet-semiconductor interfaces that produce a large number of
surface states causing strong spin relaxation and
reducing the polarization of
the injected electrons to only a few percent. Since the injected
electron crosses the metal-semiconductor interface twice in
spin-valve designs, i.e., at the source
and drain terminals of the device, the importance of the interface
problem multiplies.
Furthermore, spin relaxation in the active region of the device adds
to the complications as well.

In our study, we consider the possibilities for suppressing spin
relaxation of the conduction electrons and evaluate different
approaches. It is extremely interesting to investigate what happens
with the spin
relaxation in a 2D electron gas as the long strip of finite width is formed
using electrostatic squeesing split-gate technique or by another method.
These systems can be used to connect
active elements in integrated chip designs and can even become a part
of the active device as we continue to reduce size of the element.
Eventually we should reach the 1D limit exhibiting no spin relaxation.
This transition to the 1D behavior was recently documented in the
computer simulation of the Datta and Das spin transistor
reported by Bournel et al.\cite{bournel} that inspired our research.
We concentrate our efforts on the definition of the
cross-over regions from the point of view of the spin relaxation time and
the description of its behavior in the broad region around the
cross-over points.

The rest of this paper is organized as follows:
We start with a brief account of spin relaxation mechanisms in
semiconductors in order to identify the most relevant. We discuss its
transformation to lower-dimensional
systems, as well as different possibilities to suppress destructive spin
relaxation (Sec.~\ref{mechanisms}). The model of a narrow patterned 2D
electron gas channel is described in Sec.~\ref{model}. Results of Monte
Carlo simulation of spin relaxation in the channel are given in the
next section (Sec.~\ref{results}), along with a comprehensive discussion
of the relaxation regimes
in terms of the channel width and a value of the spin splitting of electron
subbands (Sec.~\ref{regimes}). A brief summary follows at the end.

\section{Mechanisms of spin relaxation}
\label{mechanisms}

There are several mechanisms that can
cause spin relaxation of conduction electrons\cite{oo}
(see Ref.~\onlinecite{fabian} for an up-to-date informal review):\\
{\it i}) The mechanism of D'yakonov and Perel' (DP)\cite{d'yakonov}
takes into consideration that
spin splitting of the conduction band in zinc-blende semiconductors
at finite wave vectors is equivalent to an effective magnetic field
that causes electron spin to precess. For electron experiencing
random multiple scattering events the orientation of this effective field
is random and that leads to the spin relaxation.\\
{\it ii}) Bir-Aronov-Pikus (BAP)\cite{bir} processes involve a simultaneous
flip of electron and hole spins due to electron-hole exchange
coupling.\\
{\it iii}) Spin relaxation due to momentum relaxation is possible
directly through
spin-orbit coupling [Elliot-Yafet (EY)\cite{elliott} process].\\
{\it iv}) Spin relaxation can
take place as a result of hyperfine interaction of electron spins with
magnetic momenta of lattice nuclei, the hyperfine magnetic field being
randomly changed due to the migration of electrons in the crystal.

BAP processes require a
substantial hole concentration that is not available in the
unipolar doped structures. EY processes are suppresed in 2D
environments --- target of our prime interest.
Thus, we concentrate our efforts on the DP mechanism as the most relevant
one for the case considered.

\subsection{DP mechanism and change of the dimensionality}

As a result of the relatively low
zinc-blende crystal symmetry, the effective $2\times2$
electron Hamiltonian for the conduction electrons contains spin
dependent terms that are cubic in the electron wave vector $\bbox{k}$
\begin{equation}
\label{splittingbulk}
H=\eta'[\sigma_xk_y(k_z^2-k_x^2)+\mbox{~etc~}].
\end{equation}
The constant $\eta'$ reflects the strength of the spin splitting in
the conduction band whose value is defined by the details of the
semiconductor band structure, $\sigma_i\ (i=x,y,z)$ are the Pauli
matrices, other terms in Eq.~(\ref{splittingbulk}) should be obtained by
the cyclic permutation of the indices.

As the system dimensionality is changed from the 3D to 2D by, for example,
the extremely strong spacial confinement in the third direction --- that
can be achieved in the semiconductor heterostructures \cite{SL} ---
modifications to the character of the spin relaxation occur as well.
First of all, an average wave vector in the direction of the quantum
confinement (axis $z$) is large, so the terms in the spin splitting
involving $k_z^2$ will dominate. This results in the orientation of the effective
magnetic field in the plane of the 2D electron gas ($x-y$ plane).
Still the elementary rotations around random axes, all laying in one
plane, do not commute with each other, so the electrons reaching the same
final destination by different trajectories will have different spin
orientations. For longer times, more and more distinguishable
trajectories will become possible and this will lead to a progressive reduction
of the averaged spin polarization of the electron ensemble. Since it
is rather easy to design a structure with $|k_z|>k_F$, the splitting for
a typical electron will be larger and the rate of spin relaxation will
be enhanced. Similar behavior can be naively
expected with the further reduction of the
dimensionality to the 1D case. Let the axis $x$ be along wire in what follows.
Here again the main terms will contain
spacial-confinement multiplyers $k_y^2$, $k_z^2$. The principal
difference with the 2D case
is that now all rotations are limited to a single axis direction and they 
commute with each other. Apart from the systematic rotation, spin
polarization does not disappear with time. All particles, independently
of the number and the sequence of the scattering events, that reach the
same final point B will have the same spin orientation. This statement
is relaxed if one allows intersubband scattering in the one-dimensional
system. This type of scattering becomes progressively more important for
wider and wider quantum wires with more and more subbands involved.
Thus, we will recover, as one can predict, the 2D or 3D case in
the limit of very thick quantum wires.

Actually, there are two types of terms that
appear in the effective-mass Hamiltonian for the 2D electron gas:
the bulk-asymmetry-\cite{lommer85} and structure-asymmetry-induced
(Bychkov--Rashba).\cite{bychkov,lommer88} They are of the same functional
form 
\begin{equation}
\label{splitting}
H=\eta\:\bbox{\sigma}\cdot[\bbox{k}\times\bbox{\hat{z}}]
\equiv\eta(\sigma_x k_y - \sigma_y k_x),
\end{equation}
where $\bbox{\hat{z}}$ is a unit vector in the $z$ direction.
We are not interested here in the origin of the spin splitting term in
our model problem and simply assume its presence with a wave-vector
dependence given by Eq.~(\ref{splitting}).
(see Refs.~\onlinecite{pfeffer95,pfeffer97} for the
band-structure calculation of the linear-in-$k$ spin splitting in
heterostructures simultaneously treating bulk- and structure-induced
asymmetry.)

This form of the spin Hamiltonian is equivalent to the precession
of the spin in the effective magnetic field
\begin{equation}
\label{effective}
H=\frac{\hbar}{2}\:\bbox{\sigma}\cdot\bbox{\Omega}_{\rm eff},
\mbox{~where~}\bbox{\Omega}_{\rm eff}\equiv
\eta_{\rm DP}\:\bbox{v}\times\bbox{\hat{z}}.
\end{equation}
Here the particle velocity $\bbox{v}=\hbar\bbox{k}/m^*$, and an obvious
substitution $\eta\to\eta_{\rm DP}$ is done for convenience. $\eta_{\rm
DP}$ is expressed in inversed length units. For a particle moving
ballistically the distance $1/\eta_{DP}$ spin will rotate to the angle
$\phi=1$. We remind that the quantum-mechanical description of the evolution
of the spin $1/2$ is equivalent to the consideration of the classic momentum
$\bbox{S}$ with the equation of motion
\[
\frac{d\bbox{S}}{dt}=\bbox{\Omega}_{\rm eff}\times\bbox{S}.
\]
The reciprocal effect of
electron spin on the orbital motion through spin-orbit coupling can
often be ignored due to the large electron kinetic energy in comparison
to the typical spin splittings and strong change of the momentum in
scattering events.

\subsection{Possiblities to influence spin relaxation time}

In addition to understanding of the reasons governing spin
depolarazation of carriers, we wish to consider and assess the
possibilities to actively influence these destructive processes in order
to improve parameters and gain new functionality of the future
spintronic devices. Following are the potential approaches for manipulating
spin relaxation times:\\ 
{\it i}) A rather simple observation follows directly from the essence of
the regime of motional narrowing (small elementary spin rotations during
ballistic electron flights).\cite{oo} Since $\tau_S^{-1}\sim \tau_p
\langle\Omega_{\rm eff}^2\rangle$, reduction of momentum relaxation time,
$\tau_p$, leads to the suppression of spin relaxation. On the other
hand, this will lead to increasing of broadenings as well as decoherence,
and can worsen device parameters.\\
{\it ii}) Since the bulk-asymmetry- and structure-asymmetry-induced spin
splittings are additive with the same $k$-dependence,
it is possible to tune combined spin splittings
in the conduction band to a desired value through manipulation of the
external electric field.\\
{\it iii}) Additional spin splitting, which is independent of the electron
wave vector will fix the precession axis. An evident possibility here is the
Zeeman effect. The time of spin relaxation scales in the presence of the
external magnetic field, $B$, as\cite{ivchenko}
\[
\frac1{\tau_S(B)}=\frac1{\tau_S(0)}
\frac1{1+(\Omega_{\rm L} \tau_p)^2},
\]
where $\hbar\Omega_{\rm L}=g\mu_{\rm B}B$ is a Zeeman splitting of electron
spin sublevels. This equation
suggests that for $\Omega_{\rm L} \tau_p=1$ spin relaxation time will double.\\
{\it iv}) There is a possibility of controlling the
spin relaxation of the conduction electrons by doping. The first
realization was reported in $\delta$-doped
heterostructures,\cite{wagner} and a several-orders-of-magnitude longer
spin memory has recently been observed in $n$-doped
structures.\cite{kikkawa}\\
{\it v}) When the channel width, $L$, is comparable to the magnitude of the
electron mean free path, $L_p$,
from the classical point of view, the sequential
alteration of one of the wave vector
components should effectively reduce spin relaxation (reflective
boundaries).\cite{bournel} Scattering on the boundaries (diffusive
boundaries) will decrease $\tau_p$ as well and can potentially influence
spin relaxation. Quantum mechanically, the channel narrowing
leads to the quantization of
the electron transverse motion in the strip and absence of spin relaxation
without intersubband scattering. 

The first four possibilities are considered to some extend in the
scientific literature or are just evident consequences of the relaxation
mechanisms. The fifth deserves a more thorough analysis. To check the
effect of the patterning of the 2D electron gas into a strip of some
large width $L$, we developed a simple Monte Carlo sumulator encompassing
random scattering of the particles in the channel, reflection from the
boundaries and spin rotation during free flight due to spin splitting
of the conduction band. Now we describe in more detail our working
model.

\section{Model}
\label{model}

As a model system we consider a strip of the 2D electron gas.
The third dimension (axis $z$) is quantizing and it is absolutely
irrelevant to our consideration of the particle movement in the real
space, since the intersubband gap is larger than all other energy scales
involved.\\
{\it i}) We will assume that all particles have the same velocity
$|\bbox{v}|$.\\
{\it ii}) Scattering is considered to be elastic and isotropic
in order to retain
model simplicity; the former assumption preserves velocity modulus, the
later one eliminates any correlations between directions of the
particle velocity before and after the scattering event.\\
{\it iii}) We neglect electron-electron interactions and consider all
particles to be independent.\\
{\it iv}) An assumption that all scatterers are completely uncorrelated
leads to an exponential distribution of times between any two
consecutive scattering events, their average is called the momentum
relaxation time, $\tau_p$. This time corresponds to the mean-free path
$L_p=|\bbox{v}|\tau_p$.\\
{\it v}) The problem spin Hamiltonian is given in the form of
Eq.~(\ref{splitting}) and influences only spin coordinate. We ignore the
reciprocal effect of the spin on the motion in the real space.\\
{\it vi}) The width of the channel is large, several or even tens
of mean-free paths so that it is permissible to consider classically the
electron real-space movement.\\
{\it vii}) At first we consider only reflecting boundary conditions at
the borders of the 2D strip. Reflecting channel boundaries preserve
longitudinal component of the particle velocity and change sign of the
normal component in collisions. Later, we will compare our results with
diffusive boundaries.

\subsection{Types of experiments}

For simplicity, in
all of our experiments particles will be injected into the system at some
particular point A (input terminal) at time $t=0$ with spin $\bbox{S}$.
As the particle experiences multiple scattering events a diffusive pattern of
motion is formed with, for an isotropic system, a Gaussian distribution 
\[
\Gamma(r)\sim \frac{1}{\langle r \rangle}\exp\left(-\frac{r^2}
{\langle r \rangle^2}\right),
\]
that broadens as time increases:
$\langle r \rangle \sim L_p\sqrt{t/\tau_p}$.
 
Evidently, there are multiple possibilities for experimental setups. The
definition of spin relaxation, obtained in these experiments will
vary correspondingly. Now, we consider several important possibilities:\\
{\it i}) The most informative type of experiment would be to obtain the average
spin $\langle \bbox{S} \rangle$ as a function of time $t$ at each point
B (output terminal). Results of all other experimental configurations can be
derived by partial integration of this correlation function.\\
{\it ii}) At time $t$ average $\langle \bbox{S} \rangle$ is calculated for the
whole ensemble independent of the real-space position of electrons.
Optical experiments are considered likely to deliver information of this
type, because of the limited possibilities to focus optical system and
fundamental restrictions.\\
{\it iii}) Particles, reaching output terminal are removed from the system
immediately, $\langle \bbox{S} \rangle$ is measured as a function, i.e., of
the interterminal distance. Individual particles can spend a substantial
time in the system, depending on their trajectories.
This type of experiment is the most probable variant in electric
experiments where points A and B can be identified as real device gates.
Made from ferromagnetic materials, gates can inject polarized electrons
and sense the polarization of the drain flux, delivering information
about the average spin of carriers. 

Let us now show, that our result for spin relaxation does indeed
depend on the
definition of the experiment. As a simple example we consider a pure 1D
case. From the point of view of the first and third experiments there is no spin
relaxation. There exists a systematic rotation of the spin proportional
to the distance from the injection point A. Indepent of the number of 
scattering events and individual trajectories, all particles reaching point B
will have exactly the same spin orientation, but it would be
different for different choices of point B.  
Thus, the transverse component of the spin $S_y(x)=S_y^0\cos(\eta_{\rm DP} x)$. 
For the second realization we readily conclude that the average spin
\begin{eqnarray}
\langle S_y \rangle=\int dx\: S_y(x)\Gamma(x) 
&\sim& S_y^0 \exp\left(-\frac14\eta_{\rm DP}^2\langle x
\rangle^2\right)\nonumber\\
&&\sim S_y^0 \exp\left(-\frac14\eta_{\rm DP}^2L_p^2\frac{t}{\tau_p}\right),
\nonumber
\end{eqnarray}
that is an exponential decay of spin polarization in the spatially
more and more broadening electron distribution.

Note that in the 2D case the same phenomenon of the systematic rotation
of the electron spin takes place in addition to the (2D--3D specific)
DP spin relaxation. This systematic rotation on the angle
$\bbox{\phi}=\eta_{\rm DP}\:\bbox{r}\times\bbox{\hat{z}}$ for the particle
real space transfer $\bbox{r}$ is again independent on the detailes of
individual trajectories.

\section{Results of simulation}
\label{results}

Fig.~\ref{simulation}a shows the time dependence of the average spin
polarization $2\langle S_y \rangle$ in the channel.
It is found that apart from the
systematic rotation this dependence is essentially the same for all
points inside channel that have a substantial electron occupation, so it
is possible to calculate reliably a value of spin polarization. The
calculation is performed for the D'yakonov-Perel' parameter $\eta_{\rm
DP}L_p=0.05$ and different channel widths. For this particular graph the
trajectories of $N=5\times10^3$ electrons are traced that gives a
standard deviation in the definition of the spin polarization on the
order of $\sqrt{N}\sim10^{-2}$ (throughout our investigation
$N=10^3-10^5$).

A strong dependence of the polarization decay on the channel width is
obtained. The decay is found to be approximately exponential, apart from
the small initial interval. This is the
region where the diffusive regime of electron motion and spin rotation
is not yet
established ($\sqrt{t/\tau_p}\sim1$). It overlaps or is followed by the
region where the typical trajectory of a randomly scattered particle
does not reaches channel boundaries yet. Here the relaxation should
be essentially the same as in the unpatterned 2D gas.

This generally exponential temporal behavior of spin polarization
justifies an introduction of the spin relaxation time, the dependence of
this decay time on the channel width is presented in
Fig.~\ref{simulation}b. In case of sufficiently wide channels so defined
relaxation time approaches the 2D limit of $\tau_S^{\rm 2D}$, for narrow
channels $\tau_S$ scales as $L^{-2}$.

In Fig.~\ref{simulation}c we summarize results of our simulation of
channels with a fixed width and different values of D'yakonov-Perel'
term $\eta_{\rm DP}$. As the constant $\eta_{\rm DP}$ becomes larger and
larger, the system behavior graduately goes out of the regime of
motional narrowing since elementary rotations on the elementary free
flights are not small anymore. This leads to a quick spin relaxation as
reflected by the reduction and saturation
of $\tau_S$. On the other side of this dependence
is a steep increase in the relaxation time as $\eta_{\rm DP}$ decreases.
We have found an intermediate region, where $\tau_S^{-1}$ scales as a second
power of the DP constant, that is followd by a forth-power dependence
for very small values of $\eta_{\rm DP}$.

Thus, in the well-established regime of motional narrowing
and in the limit of sufficiently narrow channels, we propose an
asymptotic formula $\tau_S\sim \eta_{\rm DP}^{-4} L^{-2}$.

To identify an effect of the particular choice of the boundary
conditions we have simulated a channel with diffusive boundaries. The
particle that reaches the channel border is scattered back into the
channel with equal probabilities for all directions of the new particle
velocity. The results of the calculation repeat closely the case of the
reflective boudaries up to the very narrow channel with widths of only
several mean free paths. No systematic deviation of the spin relaxation
time is observed for wider channels.

\section{Regimes of the DP relaxation in a channel}
\label{regimes}

Thus, we distinguish the following regimes of spin relaxation in
channels of finite width as we vary both independent parameters
$\eta_{\rm DP}$ and $L$ of the problem in consideration:

\begin{figure}
\includegraphics*[bb=89 190 540 670,height=3.2in]{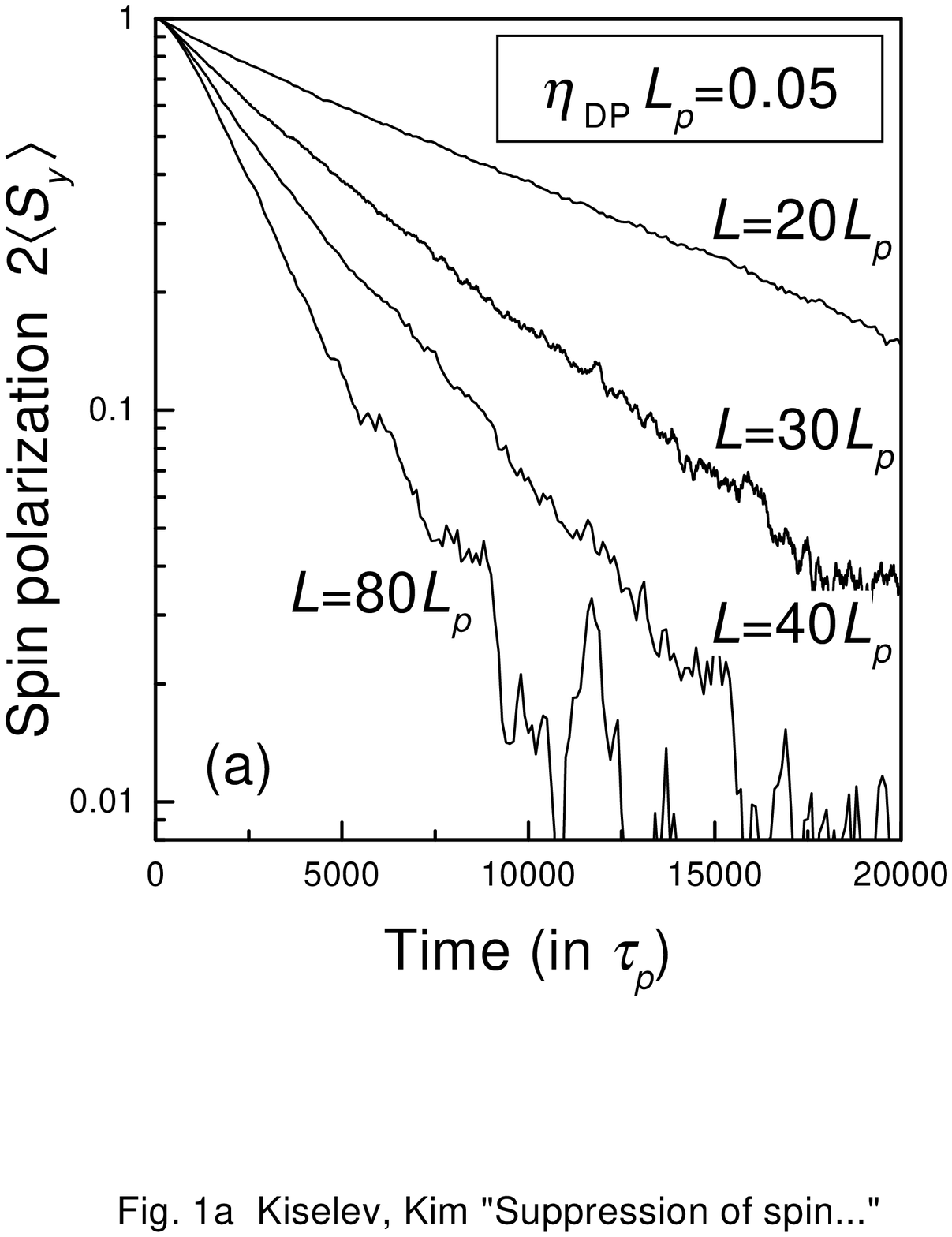}
\includegraphics*[bb=104 199 514 665,height=3.2in]{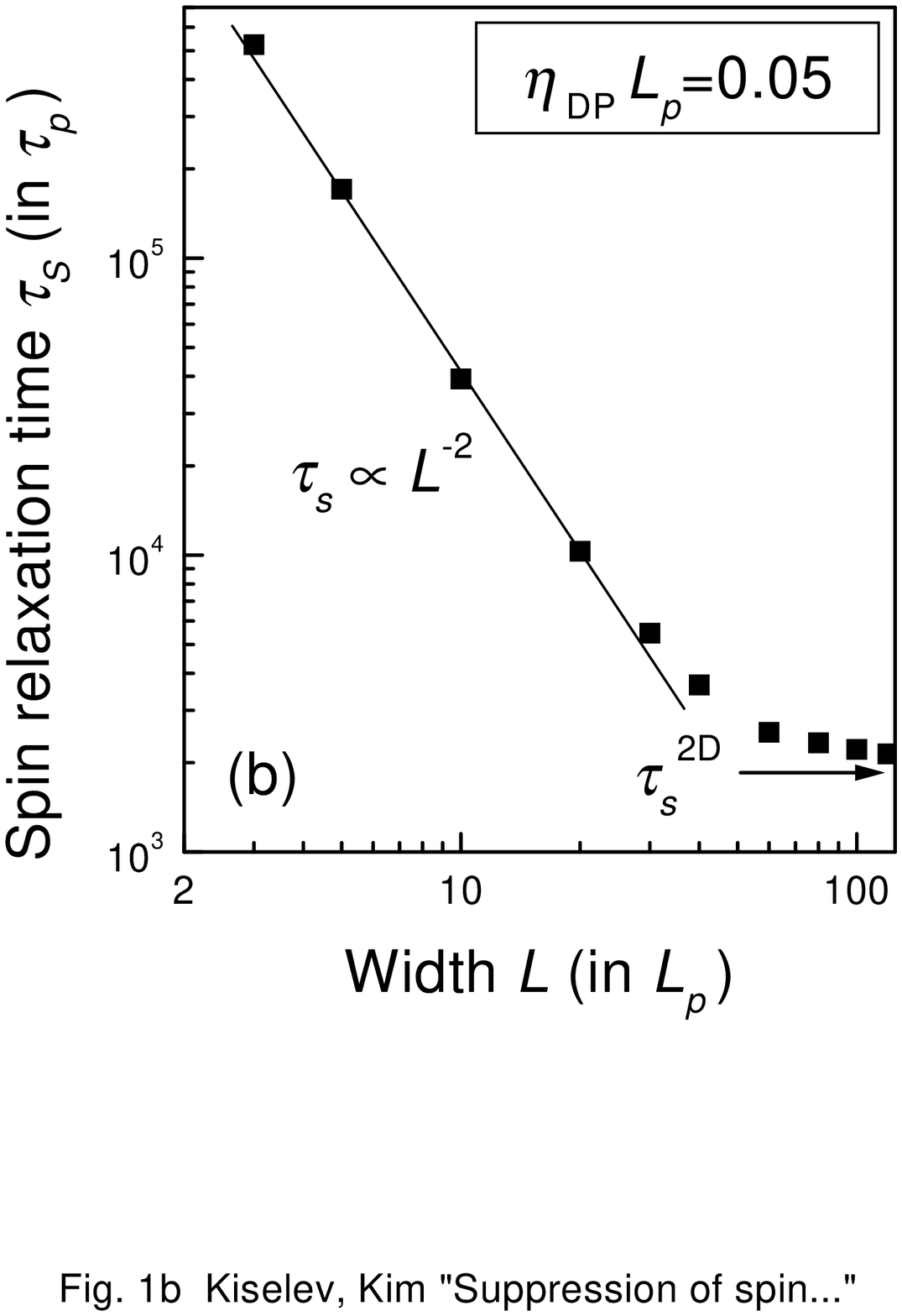}
\includegraphics*[bb=104 199 514 665,height=3.2in]{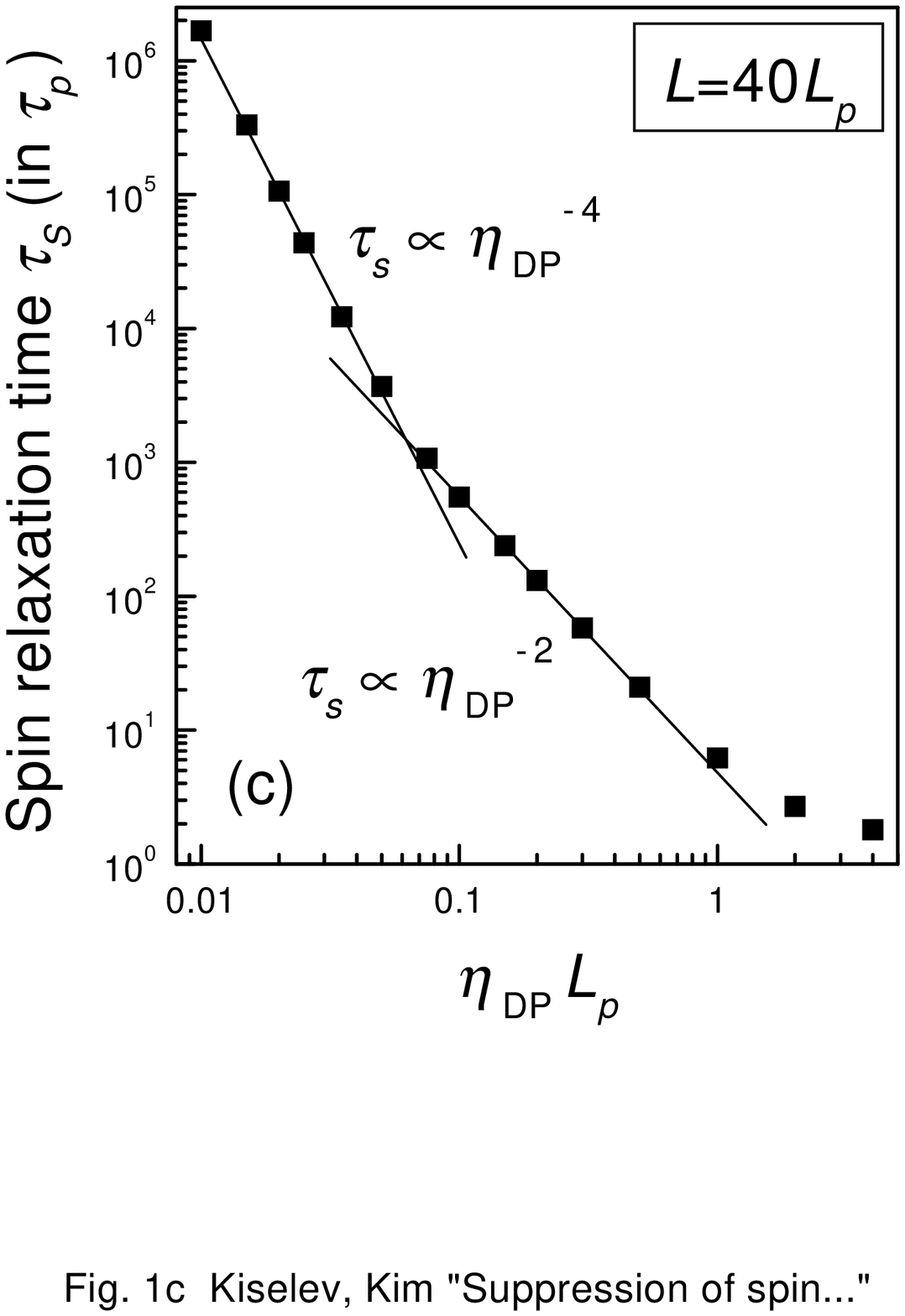}
\caption{\label{simulation}
Spin relaxation in a channel. (a) Time
dependence of the spin polarization, calculated for different channel
widths. DP constant $\eta_{\rm DP}$
is fixed to be $0.05$. The trajectories of $5\times10^3$
particles are traced that defines a standard
deviation $\sim10^{-2}$ for the calculated averages. Close-to-exponential
decays of the polarization permit to define spin relaxation time, $\tau_S$;
(b) $\tau_S$ as a function of the channel width $L$; (c) Spin relaxation
time in dependence on the DP spin splitting constant $\eta_{\rm DP}$ at fixed
channel width $L=40 L_p$.
}
\end{figure}

\begin{figure}
\includegraphics*[bb=167 340 541 728,height=3.2in]{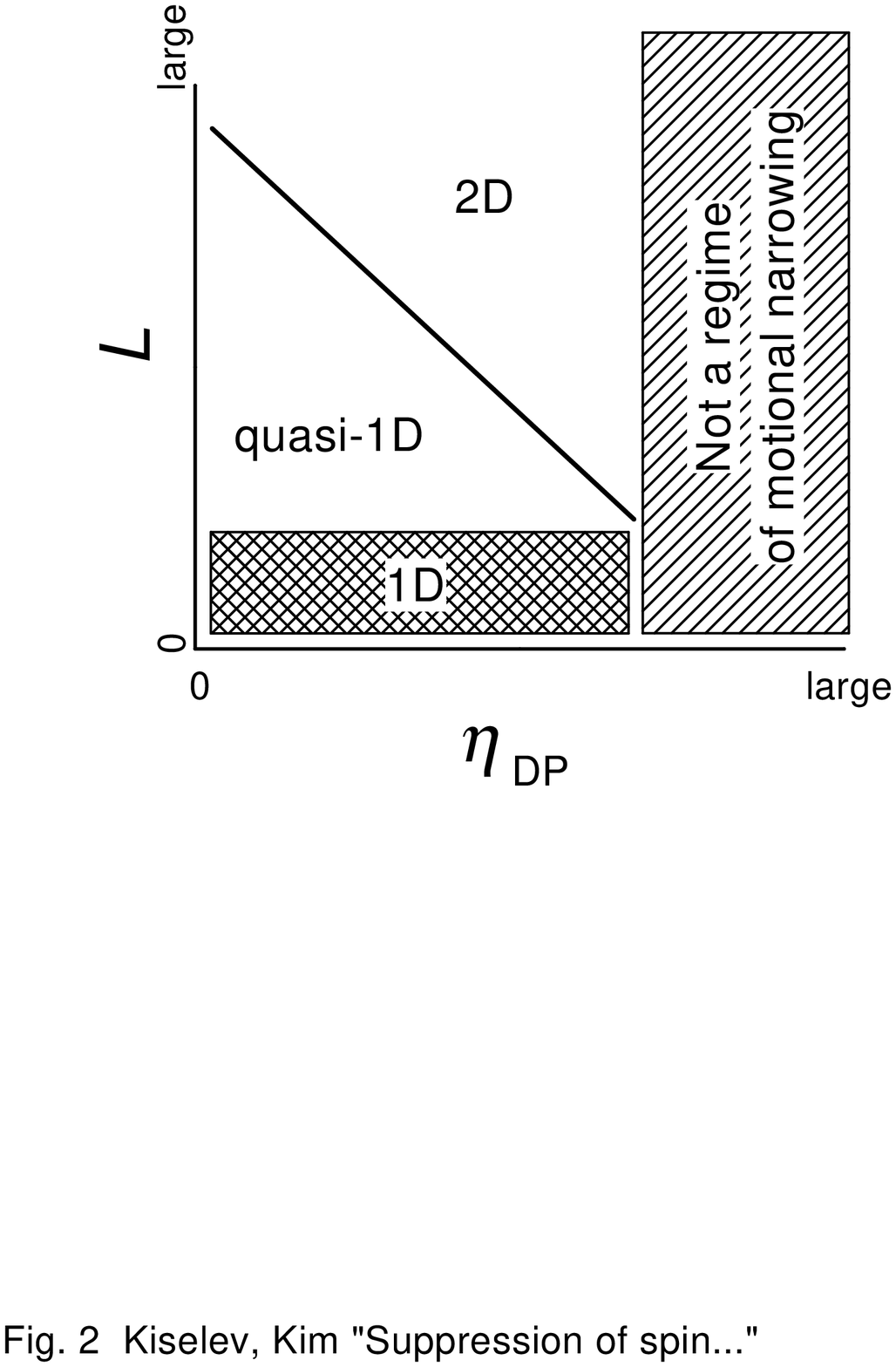}
\caption{\label{diagram}
Different regimes of the spin relaxation on the plane $(\eta_{\rm DP}, L)$ of
model parameters: $\eta_{\rm DP}L_p\protect\gtrsim 1$ ---
elemenary rotations during
free flights are not small, $\tau_{S}\sim \tau_p$;
$\eta_{\rm DP}L_p< 1$, $\eta_{\rm DP}L\protect\gtrsim 1$ ---
2D spin relaxation,
$\tau_{S}^{\rm 2D}\sim \tau_p(\eta_{\rm DP}L_p)^{-2}$;
$\eta_{\rm DP}L< 1$ --- supression of spin relaxation, quasi-1D
regime, $\tau_{S}\sim \tau_S^{\rm 2D}(\eta_{\rm DP}L)^{-2}
\sim \tau_p\:\eta_{\rm DP}^{-4}L^{-2}$;
$L\protect\lesssim L_p$ --- $L$ substitutes $L_p$, quantum mechanical
quantization in the channel.
}
\end{figure}

First of all, for very large spin splitting ($\eta_{\rm DP} L_p\gtrsim1$)
we violate a general condition for the motional narrowing regime for the
DP spin relaxation. Each elementary rotation is not small and the
information about the spin polarization is lost already after the first
random scattering event (see as well Fig.~\ref{diagram} as a guide). For
this regime $\tau_S\sim\tau_p$ and is the shortest of all regimes.

When $\eta_{\rm DP}$ is small ($\eta_{\rm DP} L_p< 1$),
we are back in the regime of motional narrowing and a well known
equation $\tau_S^{\rm 2D}\sim \tau_p (\eta_{\rm DP} L_p)^{-2}$ defines the time
of spin relaxation in the 2D system. Now we narrow the strip of the 2D
electron gas. The behavior is unchanged until $\eta_{\rm DP} L\sim 1$.
For smaller channel widths ($\eta_{\rm DP} L< 1$)
DP spin relaxation is suppresed very
effectively, with $\tau_S\sim \tau_S^{\rm 2D}(\eta_{\rm
DP} L)^{-2}$. For $L< L_p$ the channel width $L$ acts as a new mean-free
path in the system, substituting $L_p$ in equations. Actually, this
region does not satisfy our assumption about classical motion of
particles in real space; the transverse motion is quantized and this
system should be considered as a quantum wire with multiple subbands.
The cross-over points define broad regions of mixed behavior that
becomes more definitive as we move out of them. 

We conducted analysis presented here for spin relaxation in the most
convenient from our point of view units. Now, we will check that the
chosen range of parameters is well within the reasonable limits present in
the contemporary heterostructures. For the asymmetric GaAs/AlGaAs
quantum well with the electron concentration of $10^{12}$~sm$^{-2}$ in
the channel the detected experimentally and confirmed by the calculation
spin splitting in the conduction band is on the order of $0.2-0.3$~meV
at the Fermi energy (see recent paper by Wissinger et.
al\cite{wissinger} and references therein). This 2D electron
concentration corresponds to the $k_F\approx3\times10^5$~cm$^{-1}$.
Samples with $L_p\gg1/q=3\times10^{-6}$~cm are readily available in
laboratories. Furthermore, this splitting corresponds to $\eta_{\rm
DP}=6-9\times10^4$~cm$^{-1}$. This set of parameters is on the border of
the regimes of the motional narrowing and large elementary rotations.
Reducing the spin splitting (parameter $\eta_{\rm DP}$) (by, for example,
reduction of the electron concentration in the 2D electron gas and, thus,
structure asymmetry due to the internal electric field) will push parameters
into well established regime of motional narrowing, suitable for the
experimental observation of the described phenomena. Datta and Das\cite{datta}
give an estimate of $\pi/\eta_{\rm DP}=7\times10^{-5}$~cm for some
particular InGaAs/InAlAs heterostructures.

\section{Summary}

In conclusion, we have investigated spin-dependent transport in
semiconductor narrow 2D channels and explored the possibility of
suppressing spin relaxation. Our approach is based on a Monte Carlo
transport model and incorporates information on conduction band electron
spins and spin rotation mechanisms. Specifically, an ensemble of
electrons experiencing multiple scattering events is simulated
numerically to analyze the decay of electron spin polarization in channels
of finite width due to the D'yakonov-Perel' mechanism.
We have been able to identify different regimes of the spin relaxation in
the 2D channels of finite width and obtain dependencies of the spin
relaxation time on the width $L$ and DP parameter $\eta_{\rm DP}$. The
most attractive for the future spintronic applications is a regime of
the suppresed spin relaxation with the relaxation time, $\tau_S$,
scaling as $L^{-2}$.

\mbox{}\\
This work was supported, in part, by the Office of Naval Research.
We thank M. A. Stroscio for the critical reading of the manuscript.



\end{document}